\documentclass[conference]{IEEEtran}
\IEEEoverridecommandlockouts

\usepackage{algorithm}
\usepackage{algpseudocode}

\usepackage{graphicx}
\usepackage{subcaption}
\usepackage{booktabs}
\usepackage{multirow}
\usepackage{url}
\usepackage{xcolor}
\usepackage{tikz}
\usepackage{amsmath}

\usetikzlibrary{decorations.pathreplacing}
\definecolor{cRIR}{HTML}{0d366b}\definecolor{cPfx}{HTML}{2a78d6}\definecolor{cCust}{HTML}{9ec5f4}
\definecolor{cLan}{HTML}{e1e0d9}\definecolor{cIID}{HTML}{f1f0ec}\definecolor{cAcc}{HTML}{eb6834}

\def\BibTeX{{\rm B\kern-.05em{\sc i\kern-.025em b}\kern-.08em
    T\kern-.1667em\lower.7ex\hbox{E}\kern-.125emX}}
    
\begin{document}

\title{StarLoc: A Practical Approach to Encoding Geographic Locators in IPv6 for Starlink}

\author{
\IEEEauthorblockN{
Vanya Singla\IEEEauthorrefmark{1},
Mostafa Abdollahi\IEEEauthorrefmark{1},
Jinwei Zhao, and Jianping Pan
}
\IEEEauthorblockA{
Department of Computer Science, University of Victoria, BC, Canada\\
\{vanyasingla, abdolahi68, clarkzjw, pan\}@uvic.ca
}
\thanks{\textsuperscript{*} Contributed equally to this work.}
}

\maketitle

\begin{abstract}
LEO satellite networks such as Starlink provide Internet connectivity across continents, oceans, and airspace. IP geolocation providers map Starlink's IPv6 prefixes to locations mainly through its self-published GeoFeed. However, users may be far away from their assigned Points of Presence (PoPs), one label covers an entire prefix, and labels and PoP mappings can change while addresses stay the same. We characterize Starlink's IPv6 addressing using 925 historical GeoFeed snapshots (2024--2026), one year of PoP mappings, and over 7.9 million observed IPv6 addresses. Our findings reveal that customer prefixes lack clear geographic structure, prefix labels lie a median of 540\,km from their PoPs, and relabeling a single /40 can affect the reported locations of up to 65,536 possible /56 delegations. To address these challenges, we propose StarLoc, a geographic Locator--Identifier separation scheme that embeds terminal location into IPv6 addresses using a resolution-adaptive H3 Locator while preserving existing Starlink prefixes, /56 delegations, and stable Endpoint Identifiers. To balance accuracy and privacy, StarLoc reveals only cells with at least a certain number of terminals, and operators can cap resolution in sensitive regions or withhold location entirely. Unlike topology-based schemes, StarLoc encodes physical location and lets Locators follow moving users. 
\end{abstract}


\section{Introduction}\label{sec:intro}

\par Low Earth Orbit (LEO) satellite networks provide broadband Internet access across large geographic areas, including remote regions, oceans, and airspace. Starlink is one of the largest LEO networks, operating more than 11,000 satellites and serving over 13.6~million users across more than 174 countries and territories. User traffic traverses satellites, ground stations, and Starlink's backbone to an assigned home Point of Presence (PoP), where it enters the public Internet. With only about 65 publicly identified PoPs, the home PoP may be thousands of kilometers from the user (Fig.~\ref{fig:popmap}).

\par Starlink publishes an RFC~8805 GeoFeed~\cite{rfc8805}, \cite{starlink_geoip_feed} that labels each prefix with a country, region, and ``city'', and a PoP feed~\cite{starlink_pops} that maps prefixes to PoPs. IP geolocation providers use such feeds as a primary input for content localization, DNS and CDN server selection, fraud detection, and access policies~\cite{iab_ipgeo_report}. Our earlier case study~\cite{pan2025geofeed} and recent work~\cite{simon2026starlink} show that these feeds change often. For IPv4, this churn stems from address scarcity: Starlink keeps acquiring new /24 blocks and moving its limited ones between PoPs and countries, with 944 IPv4 relabels from January 2024 to October 2026 and 497 IPv4 PoP changes in one year. IPv6 address space is abundant, and we observe only 146 IPv6 relabels and 81 PoP changes, of which 22 accompany newly opened PoPs and 27 belong to three bulk consolidations. Once deployed, IPv6 prefixes rarely move.

\par For Starlink's IPv6, the real problem is granularity and accuracy. To quantify it, we combine 925 GeoFeed snapshots (2024--2026), one year of PoP mappings, and weekly IPv6 measurements (Oct.\ 2025--Oct.\ 2026) covering over 7.9 million observed active router addresses~\cite{wang2026large}. We find that (i)~one city label covers a whole prefix, i.e., up to 65,536 customer /56s in a /40. Outside the US (63 IPv6 city labels), Canada (12), and Australia (6), 213 of 221 countries and territories have a single IPv6 city label, usually the capital or largest city. (ii)~Labels are far away from where traffic exits: a prefix's city lies a median of 540\,km from its PoP, e.g., users in Zimbabwe were initially served through Frankfurt~\cite{bose2026investigating}. (iii)~Customer /56s are allocated without regard to location, so finer labels cannot be derived from current addresses. As geolocation providers inherit these labels, users receive content for the wrong region, distant DNS and CDN servers~\cite{bose2026investigating}, and the wrong access policies.

\par In this paper, we propose StarLoc, which encodes the user terminal's location, as an H3 cell~\cite{h3}, into the operator-controlled bits between a published prefix and the customer's /56 local area network (LAN) prefix. StarLoc preserves Starlink's existing prefixes and /56 delegation structure while separating a geographic Locator from a stable Endpoint Identifier, so the Locator can follow moving terminals. Per-prefix code tables describe the Locators, and their rows are CIDR blocks that can also be published as RFC~8805 entries. For privacy, StarLoc discloses at most a cell with at least $k$ terminals and a minimum area, and can let operators cap the resolution of sensitive regions regardless of demand and provide non-geographic Locators for opt-out and anycast addresses. Unlike LISP~\cite{rfc9300}, HIP~\cite{rfc7401}, and ILNP~\cite{rfc6741}, which represent network topology, StarLoc encodes physical location. In eight PoP regions on four continents, StarLoc reduces mean positioning error from 275--2,203\,km with PoP-based geolocation to about 6\,km, i.e., by up to 141$\times$, using only 12--14 Locator bits.

\par The remainder of the paper is organized as follows: Sec.~\ref{sec:starlink} gives background, Sec.~\ref{sec:analysis} characterizes Starlink's IPv6 addressing, Sec.~\ref{sec:design} presents StarLoc, Sec.~\ref{sec:eval} evaluates it, and Sec.~\ref{sec:concl} concludes the paper.

\section{Background and Related Work}\label{sec:starlink}

\subsection{Starlink IPv6 Addressing}

\par Starlink supports native IPv6 on all routers, kit versions, and service plans, whereas its default IPv4 service uses carrier-grade NAT~\cite{starlink_ipv6_help}. Each terminal receives one /64 for its WAN via SLAAC and one /56 for its LAN via DHCPv6 prefix delegation (DHCPv6-PD)~\cite{starlink_ipv6_help}, i.e., 256 /64 subnets, in line with RFC~6177~\cite{rfc6177}. A reservation system keeps the prefix across reboots, but relocation or software updates may change it~\cite{starlink_ipv6_help}. Customer /56s are drawn from five RIR blocks: \texttt{2605:59c0::/28} (ARIN), \texttt{2406:2d40::/32} and \texttt{2401:e320::/32} (APNIC), \texttt{2803:9810::/32} (LACNIC), and \texttt{2a0d:3340::/29} (RIPE NCC). Starlink divides them into /38--/48 prefixes, each with one GeoFeed label and one PoP (Fig.~\ref{fig:ipv6_structure}); a /40, for example, leaves 16 bits to select up to 65,536 customer /56s.

\begin{figure}[!t]
    \centering
    \begin{tikzpicture}[x=0.0655cm,y=1cm,font=\scriptsize]
  \fill[cRIR] (0,0) rectangle (32,0.42);
  \fill[cPfx] (32,0) rectangle (40,0.42);
  \fill[cCust] (40,0) rectangle (56,0.42);
  \fill[cLan] (56,0) rectangle (64,0.42);
  \fill[cIID] (64,0) rectangle (128,0.42);
  \draw[white,line width=0.8pt] (32,0)--(32,0.42) (40,0)--(40,0.42) (56,0)--(56,0.42) (64,0)--(64,0.42);
  \node[white] at (16,0.21) {RIR block};
  \node[white,font=\bfseries\scriptsize] at (36,0.21) {1};
  \node[font=\bfseries\scriptsize] at (48,0.21) {2};
  \node[font=\bfseries\scriptsize] at (60,0.21) {3};
  \node at (96,0.21) {Interface Identifier};
  \foreach \b in {0,56,64,128} {\node[above,inner sep=1pt] at (\b,0.42) {\tiny\b};}
  \draw[decorate,decoration={brace,mirror,amplitude=3pt}] (0.5,-0.04)--(39.5,-0.04) node[midway,below=2pt] {published prefix};
  \draw[decorate,decoration={brace,mirror,amplitude=3pt}] (40.5,-0.04)--(55.5,-0.04) node[midway,below=2pt] {slot};
  \draw[decorate,decoration={brace,mirror,amplitude=3pt}] (56.5,-0.04)--(127.5,-0.04) node[midway,below=2pt] {inside the customer's /56};
  \node[anchor=north west,text width=8.4cm,align=left,inner sep=0pt] at (0,-0.48) {\hyphenpenalty=10000\exhyphenpenalty=10000%
    \textbf{1}\,Published prefix, /38--/48 (828 prefixes, of which 82\% are /40s): \textcolor{cAcc}{\textbf{one city label}} (GeoFeed) and \textcolor{cAcc}{\textbf{one PoP}} (PoP feed).\\
    \textbf{2}\,Customer slot, 8--18 bits (16 bits for a /40), chosen by DHCPv6-PD: \textcolor{cAcc}{\textbf{no geographic meaning}}.\\
    \textbf{3}\,LAN subnet (8 bits, 256 /64s), then the 64-bit IID set by hosts.\\[1pt]
    e.g., \texttt{2605:59c0:5400::/40} $\rightarrow$ ``CA, Calgary'', \texttt{clgycan1}, for up to $2^{16}$ customers, each having a /56.};
\end{tikzpicture}
    \caption{Starlink IPv6 address structure.}
    \label{fig:ipv6_structure}
\end{figure}

\par \textbf{Why the /56 matters.} The /56 is the unit of a Starlink subscriber. Bits 0--55 are set by the operator, whereas bits 56--127 belong to the customer's LAN and hosts, so location information that the customer cannot alter must sit above the /56 boundary. The /56 also makes Starlink measurable from outside: because most terminals lack public IPv4 addresses, Wang et al.~\cite{wang2026large} probe the GeoFeed prefixes over IPv6 and discover millions of active user routers, whose addresses we map to their /56 positions (Sec.~\ref{sec:analysis}).

\subsection{Locator--Identifier Separation and Geographic Addressing}

\par LISP~\cite{rfc9300} maps endpoint identifiers to routing locators through a mapping system, HIP~\cite{rfc7401} binds cryptographic host identities to IP locators, and ILNP~\cite{rfc6741} splits an IPv6 address into a 64-bit Locator and a 64-bit Node Identifier. These schemes support mobility and multihoming, but their Locators name points of network attachment rather than physical positions. Geographic addressing has also been proposed: Hain derives provider-independent IPv6 prefixes from interleaved latitude and longitude bits~\cite{hain_geo_addr}, and SpaceCore uses geospatial addressing and service areas to make LEO mobile core functions stateless~\cite{li2022spacecore}. Self-published GeoFeeds~\cite{rfc8805} are a primary input to geolocation databases. Their gaps were discussed at the IAB workshop~\cite{iab_ipgeo_report} and measured for Starlink~\cite{pan2025geofeed,simon2026starlink}. StarLoc instead embeds a privacy-bounded, resolution-adaptive H3 cell in the operator-controlled bits of an existing ISP addressing plan, without new routing or mapping infrastructure.

\section{Measurement and Analysis}\label{sec:analysis}

We combine three datasets: \textbf{(i)} 925 GeoFeed snapshots (January 2024--October 2026), archived by us~\cite{zhao_starlink_geoip_data}, with country, region, and city labels; \textbf{(ii)} 302 snapshots of the PoP feed (October 2025--October 2026, since its availability); and \textbf{(iii)} weekly IPv6 measurements~\cite{wang2026large} (October 2025--October 2026) with over 7.9 million observed active addresses, each mapped to its /56. We also use a snapshot (October 23, 2026) of the BGP announcements of Starlink's AS14593 and its Indonesian AS45700 from BGP.tools~\cite{zhao_starlink_geoip_data}.

\subsection{From RIR Blocks to Active Customers}







\begin{table}[!t]
\centering
\caption{Starlink customer IPv6 space in millions of /56s: RIR allocation, BGP announcements, GeoFeed coverage, and active /56s (October 2026).}
\label{tab:funnel}
\vspace{-1mm}
\scriptsize
\setlength{\tabcolsep}{3.2pt}
\begin{tabular}{@{}l l r r r r@{}}
\toprule
\textbf{RIR block} & \textbf{RIR} & \textbf{Allocated} & \textbf{BGP} & \textbf{GeoFeed} & \textbf{Active} \\
\midrule
\texttt{2605:59c0::/28} & ARIN & 268.4 & 33.2 & 32.9 & 4.2\\
\texttt{2406:2d40::/32} & APNIC & 16.8 & 10.2 & 10.4 & 0.8\\
\texttt{2401:e320::/32} & APNIC & 16.8 & 1.0 & 1.0 & 0.1\\
\texttt{2803:9810::/32} & LACNIC & 16.8 & 8.7 & 8.7 & 1.9\\
\texttt{2a0d:3340::/29} & RIPE & 134.2 & 10.4 & 10.7 & 1.0\\
\midrule
Total & & 453.0 & 63.4 & 63.8 & $\sim$8.0 \\
\bottomrule
\end{tabular}
\vspace{-3mm}
\end{table}

Table~\ref{tab:funnel} follows Starlink's IPv6 space from allocation to use. The five RIR blocks hold 453.0M possible customer /56s. BGP announces 63.4M of them (14.0\%) in 1,069 prefixes, and the GeoFeed allocates 63.8M (14.1\%) in 828 prefixes, of which 678 (82\%) are /40s. BGP and the GeoFeed largely agree: 810 GeoFeed prefixes are announced, 18 (1.5M /56s) are published before being routed, and 43 announced blocks (1.1M /56s) have no GeoFeed entry. Usage differs by block: the APNIC \texttt{2406:2d40::/32} and the LACNIC block are 61\% and 52\% announced, respectively, whereas the ARIN /28 is only 12\% announced. Only 7.9M /56s ($\sim$ 1.8\% of the GeoFeed capacity) were active in our October 18, 2026 snapshot. Address space is therefore not a constraint: about 390M unannounced /56s remain, enough to renumber customers under a new addressing plan without new RIR allocations.

\begin{figure}[!t]
  \centering
  \includegraphics[width=\columnwidth]{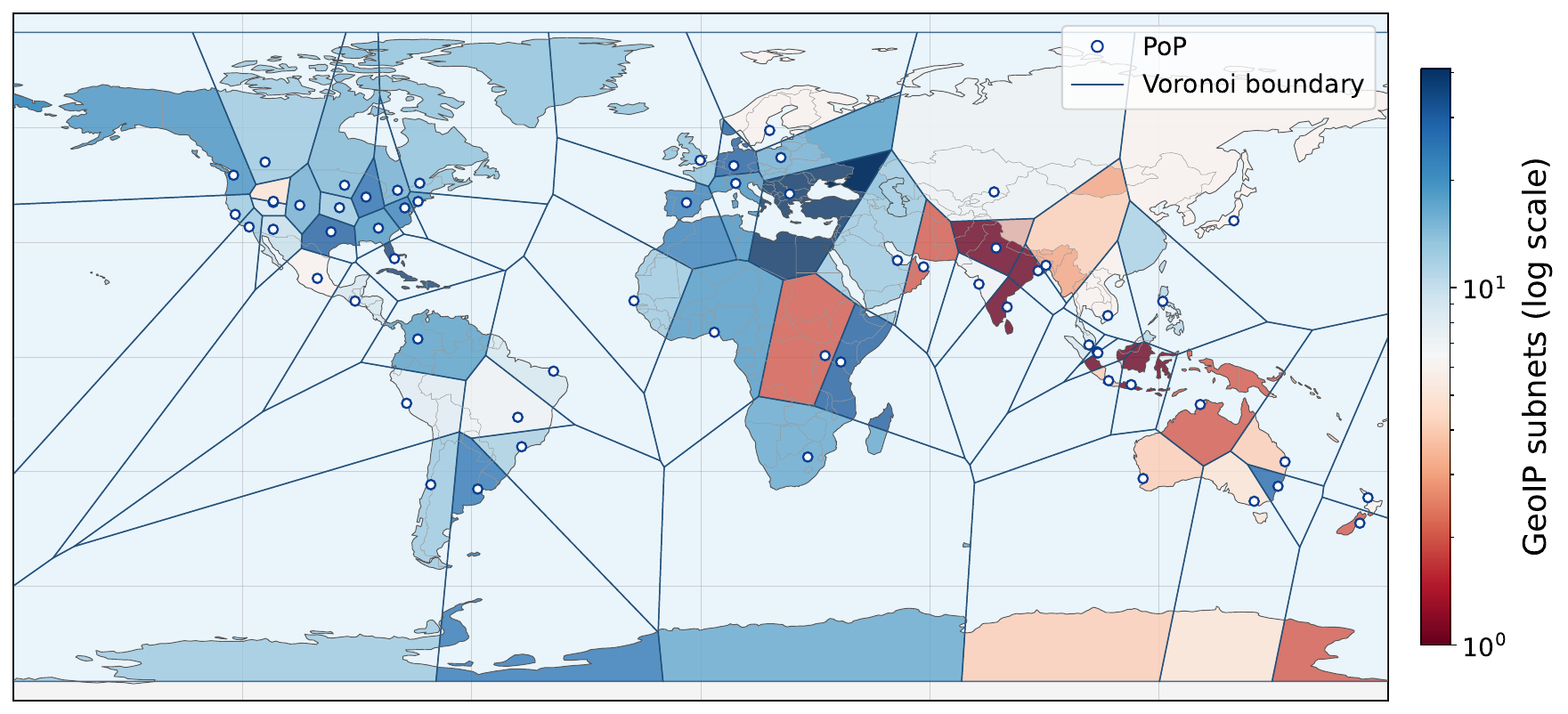}
  \vspace{-5mm}
  \caption{Published IPv6 prefixes per PoP, shown on the Voronoi regions of the PoPs (log scale, Oct.\ 2026).}
  \label{fig:popmap}
  \vspace{-3mm}
\end{figure}

Fig.~\ref{fig:popmap} maps the 749 IPv6 prefixes that have a PoP onto distance-based Voronoi regions of the 65 PoPs, which approximate
Starlink's PoP service boundaries~\cite{zhao2026aviation}. The other 79 prefixes have no PoP mapping, including all 23 subnets of aviation ``anycast users''~\cite{owens2025starlink,zhao2026aviation}. The median PoP serves 10 prefixes and Sofia does the most (41), whereas several recently opened PoPs in India, Southeast Asia, and Uganda serve only one, two of which are not yet announced in BGP.  Every region nevertheless spans thousands of kilometers, often including oceans and polar areas, so a PoP locates its users as coarsely as a city label does. StarLoc therefore uses H3 cells rather than PoP regions as its unit of location for consistency and comparability (Sec.~\ref{sec:design}).

\subsection{Allocation Patterns Inside a Prefix}

\begin{figure}[!t]
  \centering
  \includegraphics[width=\columnwidth]{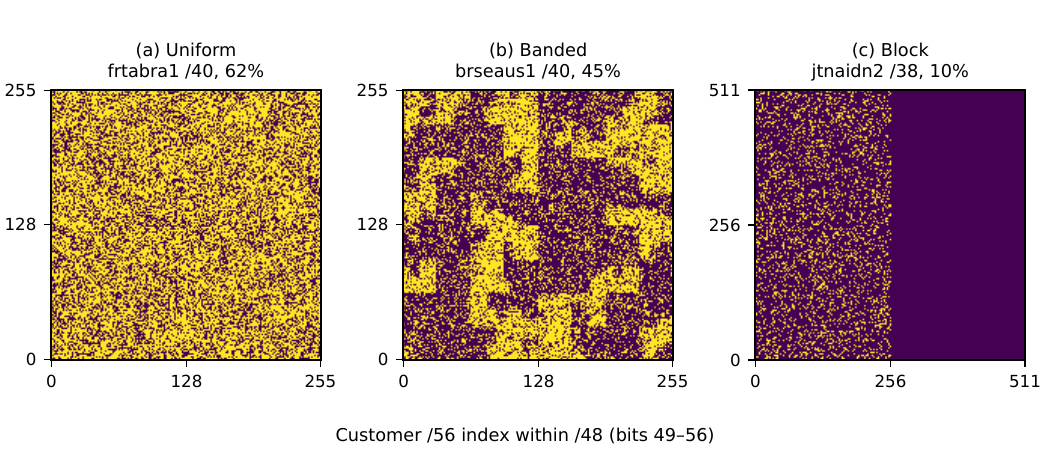}
  \vspace{-5mm}
  \caption{Three allocation patterns of active customer /56s inside a published prefix (yellow = observed). Snapshots: (a) 2025-10-06, (b) 2026-04-05, (c) 2026-10-11.}
  \label{fig:patterns}
  \vspace{-2mm}
\end{figure}

How does Starlink place customer /56s inside a published prefix? We map each prefix to a grid with one row per /48 and one column per /56 within it (bits 49--56), and mark each observed /56. Fig.~\ref{fig:patterns} shows the three patterns we find. \emph{Uniform} prefixes (Fig.~\ref{fig:patterns}a) are filled evenly, consistent with random selection from a pool. \emph{Banded} prefixes (Fig.~\ref{fig:patterns}b) concentrate customers in selected /48 rows. \emph{Block-concentrated} prefixes (Fig.~\ref{fig:patterns}c) use only part of a larger prefix, for example the lower half of a /38. Within one PoP, prefixes can follow different patterns, two large prefixes of the Calgary PoP are 8.5\% and 37.3\% occupied, one sparse everywhere and one dense in two blocks, suggesting prefix-specific allocation pools or batch migrations.

\begin{figure}[!t]
  \centering
  \includegraphics[width=0.8\columnwidth]{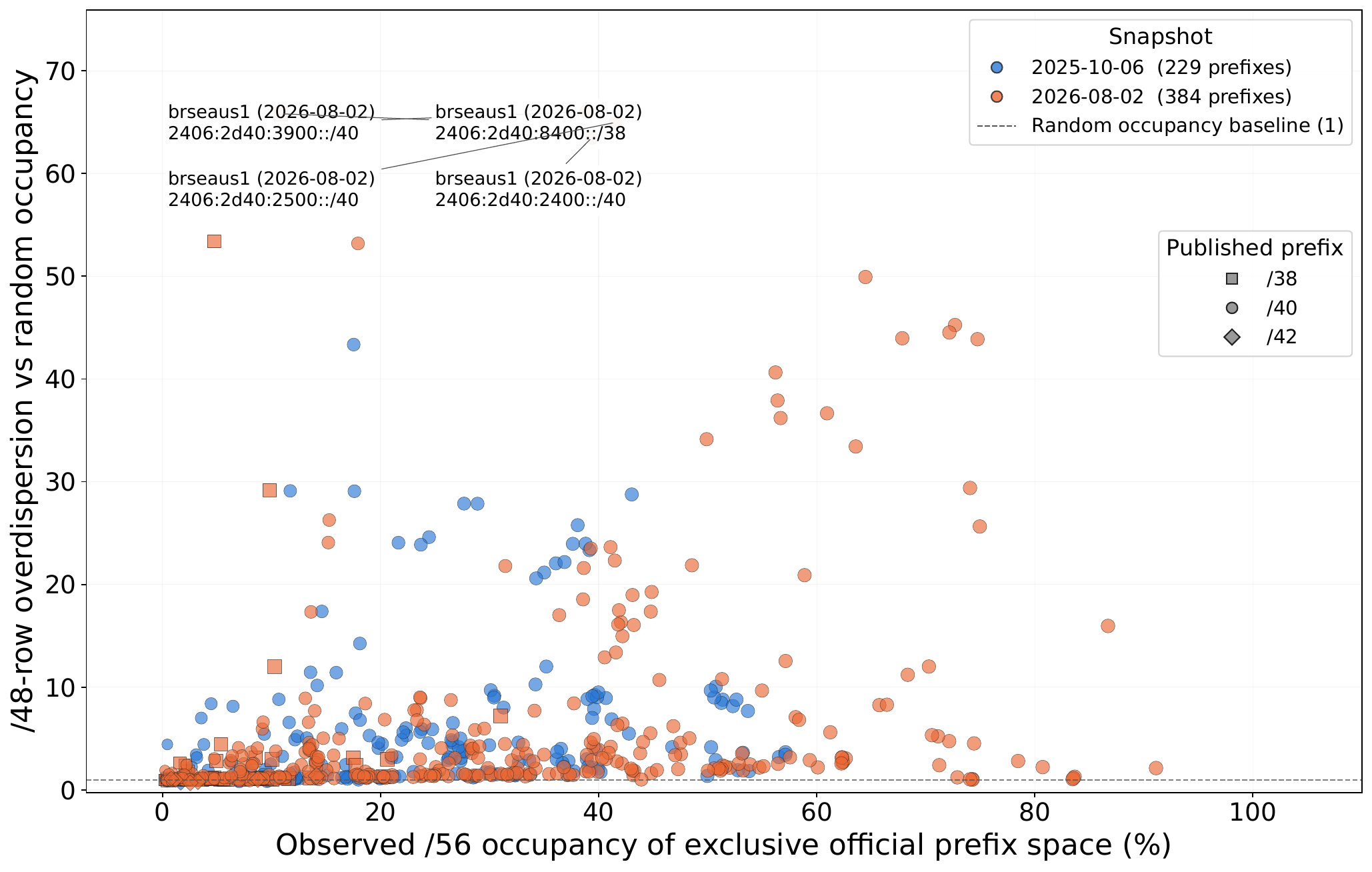}
  \vspace{-1mm}
  \caption{Occupancy vs.\ /48-level overdispersion per published prefix, Oct.\ 2025 (229 prefixes) and Aug.\ 2026 (384 prefixes).}
  \label{fig:occupancy}
  \vspace{-3mm}
\end{figure}

To compare all prefixes, Fig.~\ref{fig:occupancy} plots each prefix's occupancy against the overdispersion of its per-/48 occupancy relative to random placement (1 = random). Between Oct.\ 2025 and Aug.\ 2026, prefixes filled substantially: most were below 55\% occupied in 2025, and many reach 60--90\% in 2026. Most prefixes stay close to random, and those that cluster do so only temporarily (Sec.~\ref{sec:temporal}). Regardless of the pattern, the choice of /56 within a prefix carries no geographic meaning. How Starlink partitions its DHCPv6-PD pools remains an open question for the operator. Even at the prefix level, numbering ignores geography. Of 608 numerically adjacent /40 pairs, 65\% are labeled in different countries and 46\% are served by different PoPs, with a median distance of 1,265\,km between their labels.

\subsection{Temporal Changes: Case Studies}\label{sec:temporal}

\begin{figure}[!t]
  \centering
  \includegraphics[width=0.8\columnwidth]{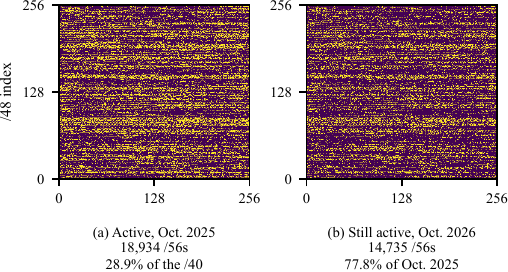}
  \vspace{-2mm}
  \caption{Stability of customer /56s in a Calgary /40 (\texttt{2605:59c0:5700::/40}). (a) Active /56s on 2025-10-06. (b) The subset still active on 2026-10-18.}
  \label{fig:stability}
  \vspace{-3mm}
\end{figure}

\textbf{Clustering comes and goes.} The Montréal and Dallas prefixes that were clustered in 2025 returned to near random by 2026, whereas Brisbane, Christchurch, and new US regional prefixes became strongly clustered (Fig.~\ref{fig:occupancy}). The Brisbane prefix in Fig.~\ref{fig:patterns}(b) was uniform in October 2025 and banded by April 2026. Bands may form when customers are moved into a prefix in batches and dissolve as it fills, although the measurements alone cannot yet establish the cause.

\textbf{Persistence in Calgary.} Fig.~\ref{fig:stability} follows \texttt{2605:59c0:5700::/40}, which had 18,934 active /56s on October 6, 2025. Of these, 14,735 (77.8\%) were still active on October 18, 2026, and 12,198 (64.4\%) were observed in every one of the eight snapshots in between. Across the most stable /40s, this persistence ranges from 52\% to 67\%. Repeated observation of a /56 does not guarantee that it stayed with the same customer, and non-continuous measurements may underestimate persistence.

\textbf{Prefix lifecycle.} Averaged over the $n$ prefixes in which
each transition was observed, newly published prefixes were first
observed active after 70 days ($n=72$), reassigned prefixes moved to
another PoP 262 days after publication ($n=54$), and removed prefixes
had been published for 550 days ($n=484$).

\subsection{GeoFeed Labels: Granularity, Accuracy, and Churn}

\begin{figure}[!t]
  \centering
  \includegraphics[width=\columnwidth]{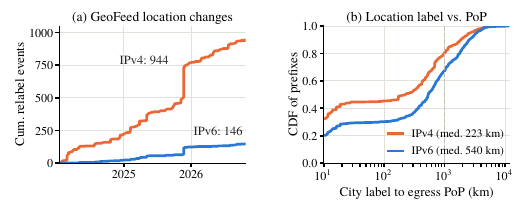}
  \vspace{-5mm}
  \caption{(a) Cumulative GeoFeed relabels (same prefix, new location), Jan.\ 2024--Oct.\ 2026. (b) Distance from a prefix's GeoFeed city to its egress PoP (Oct.\ 2026).}
  \label{fig:dynamics}
  \vspace{-3mm}
\end{figure}

\textbf{Coarse labels.} The GeoFeed assigns one label to each published prefix, so customers anywhere in a prefix's service area share one city. Labels are dense only in the US (63 IPv6 city labels), Canada (12), and Australia (6). The other 221 countries and territories have 235 labels in total, and 213 of them have exactly one, even Brazil and India have only five and four. These single-label countries hold 52\% of the published /56 capacity, and no IPv6 prefix is labeled in Yukon or the Northwest Territories in Canada.

\textbf{Labels far away from PoPs.} A prefix's label lies a median of 540\,km from its egress PoP for IPv6, versus 225\,km for IPv4 (Fig.~\ref{fig:dynamics}b), so a label does not even indicate where traffic leaves Starlink's network.

\textbf{Relabeling and PoP changes.} Between January 2024 and October 2026, we identified 944 IPv4 and 146 IPv6 relabeling events (Fig.~\ref{fig:dynamics}a), 50 of which changed an IPv6 prefix's country. Although IPv6 relabels are 6.5$\times$ rarer, together they relabeled about 6.9M possible /56s, versus 183K IPv4 addresses. In its first year, the PoP feed recorded 497 PoP changes for IPv4 /24s but only 81 for IPv6 prefixes. Of these 81, 10 moved prefixes out of a placeholder PoP and 12 into one of the 17 PoPs that first appeared during that year. Another 27 belong to three bulk consolidations that moved Pacific-island, Caribbean, and South Asian prefixes to Sydney, Miami, and Singapore, respectively, while their labels stayed in their own capitals. Previous research found that a PoP reassignment could appear in routing at least 48 hours before the corresponding feed update~\cite{simon2026starlink}.

\par \textbf{Implications.} Starlink has ample IPv6 space, but neither its numbering nor its GeoFeed conveys where customers are: /56s are placed without regard to location, labels are coarse and far away from both users and PoPs, and 79 prefixes lack PoP mappings. To address these challenges, a geographic addressing system must \textbf{R1} preserve existing RIR blocks, published prefixes, /56 delegations, and /64 subnets; \textbf{R2} encode terminal location at privacy-tunable granularity independently of PoPs; \textbf{R3} enable location decoding through per-prefix rules; and \textbf{R4} separate a mobility-aware Locator from a stable Endpoint Identifier. We introduce StarLoc (Sec.~\ref{sec:design}) to satisfy these four requirements.

\section{StarLoc Design}\label{sec:design}

\subsection{System Model and Address Structure}
\label{sec:model}

In StarLoc design, locations are H3 cells. Each cell $c$ has resolution $r_c$, edge length $a_{r_c}$, area $A_c$, and demand $d_c$
(its number of terminals). Each terminal has a service class $\kappa$ (fixed, roaming, maritime, or aviation) that caps its resolution
at $r_{\max}(\kappa)$.  StarLoc assigns each service region (initially one PoP's region) a routable IPv6 service-region prefix of $P$ bits (e.g., /35), allowing PoP reassignment without renumbering. The remaining $56-P$ bits contain a geographic Locator ($L$ bits) followed by a customer slot ($S$ bits), as shown in Fig.~\ref{fig:starloc_addr}. The Locator comprises a 3-bit H3 resolution field ($r=1$--$5$ and code 0 for a withheld Locator, with codes 6 and 7 reserved) and a $b$-bit H3 cell index identifying a compacted cell through a per-prefix mapping (Sec.~\ref{sec:compaction}). Thus, $L=3+b$ and $S=56-P-L$. With $b=9$, the Locator occupies 12 bits, and each cell index supports $2^S$ customer /56 delegations. Each /56 retains 8 bits for LAN subnets and a 64-bit Interface Identifier (IID).

\begin{figure}[!t]
\centering
\begin{tikzpicture}[x=0.0655cm,y=1cm,font=\scriptsize]
  \fill[cPfx] (0,0) rectangle (35,0.42);
  \fill[cAcc] (35,0) rectangle (38,0.42);
  \fill[cAcc!45] (38,0) rectangle (47,0.42);
  \fill[cCust] (47,0) rectangle (56,0.42);
  \fill[cLan] (56,0) rectangle (64,0.42);
  \fill[cIID] (64,0) rectangle (128,0.42);
  \draw[white,line width=0.8pt] (35,0)--(35,0.42) (38,0)--(38,0.42) (47,0)--(47,0.42) (56,0)--(56,0.42) (64,0)--(64,0.42);
  \node[white] at (17.5,0.21) {IP prefix ($P$ bits)};
  \node[white] at (36.5,0.21) {$r$};
  \node at (42.5,0.21) {index};
  \node at (51.5,0.21) {$S$};
  \node at (60,0.21) {LAN};
  \node at (96,0.21) {Interface Identifier (IID)};
  \foreach \b/\t in {0/0,35/$P$,56/56,64/64,128/128} {\node[above,inner sep=1pt] at (\b,0.42) {\tiny\t};}
  \draw[decorate,decoration={brace,mirror,amplitude=3pt}] (0.3,-0.04)--(34.7,-0.04) node[midway,below=2pt] {Starlink prefix};
  \draw[decorate,decoration={brace,mirror,amplitude=3pt}] (35.3,-0.04)--(46.7,-0.04) node[midway,below=11pt,xshift=4pt] {Locator ($L{=}3{+}9$)};
  \draw[decorate,decoration={brace,mirror,amplitude=3pt}] (47.3,-0.04)--(55.7,-0.04) node[midway,below=2pt,xshift=2pt] {$S$};
  \draw[decorate,decoration={brace,mirror,amplitude=3pt}] (64.5,-0.04)--(127.5,-0.04) node[midway,below=2pt] {host-generated, changes with the prefix};
\end{tikzpicture}\\[2pt]
\scriptsize
\setlength{\tabcolsep}{3.5pt}
\begin{tabular}{@{}c c c c r@{}}
\toprule
$P$ & res.\ (bits) & index (bits) & $S=56{-}P{-}12$ & \textbf{/56s per index} \\
\midrule
32 & 3 & 9 & 12 & 4,096 \\
35 & 3 & 9 & 9 & 512 \\
38 & 3 & 9 & 6 & 64 \\
\bottomrule
\end{tabular}\\[3pt]
Example (Calgary cell \texttt{8512ccd7fffffff}, $r{=}5$, index 14, slot 23):\\
\texttt{2605:59c1:8000::/35}\,|\,\texttt{101}\,|\,\texttt{000001110}\,|\,\texttt{000010111} $\rightarrow$ \texttt{2605:59c1:941c:1700::/56}
\vspace{-1mm}
\caption{StarLoc address structure ($P{+}L{+}S{=}56$).}
\label{fig:starloc_addr}
\vspace{-3mm}
\end{figure}

\par {\bf Region-Based Geolocation Accuracy and Privacy Constraints:} StarLoc balances geolocation accuracy, privacy, address capacity, and mobility through adaptive H3 resolution. Finer cells improve positioning accuracy, with expected error $E_R(r)$, the mean terminal-to-cell-center distance, $\approx0.61a_r$ (6--295\,km for $r=5$--$1$), but may expose precise locations or consume excessive address space. To preserve privacy, each cell $c$ must satisfy $A_c\geq A_{\min}$ and $d_c\geq k$, where $A_c$ is its area and $d_c$ its terminal demand. Address capacity limits the number of allocated indices at each resolution to $2^b$, while service-specific bounds $r_c\leq r_{\max}(\kappa)$ restrict location precision for mobile terminals. Demand-aware compaction (Sec.~\ref{sec:compaction}) selects resolutions and indices under these constraints.

\subsection{Demand-Aware Cell Compaction and Indexing}
\label{sec:compaction}

Each index provides $2^S$ customer /56 slots. Assigning one index per cell wastes capacity in sparse regions while limiting busy cells. Therefore, sparse sibling cells are merged into their H3 parent, increasing their combined demand $d_c$ and supporting $k$-anonymity, while dense cells retain finer resolution. Cells with high demand receive multiple consecutive indices:
\begin{equation}
n_c = 2^{\lceil\log_2\lceil(1+\gamma)\,d_c/2^{S}\rceil\rceil},
\label{eq:indices}
\end{equation}
where $\gamma$ represents growth headroom, $\theta$ is the merge threshold as a fraction of one index, and the power-of-two allocation enables prefix-aligned address blocks. Algorithm~\ref{alg:compact} applies this procedure by merging cells when demand or mobility constraints require it and repeating compaction when index capacity is exceeded. The algorithm is deterministic, with ties broken by H3 ID, allowing the resulting code table to be reproduced from the same inputs.

\textbf{Sensitive and unserved regions.} High demand does not imply high precision: for areas such as Ukraine or Iran, the operator sets a cap $r_{\mathrm{pol}}(c)$ (line~5), so demand only adds indices $n_c$ at the capped resolution, and code~0 withholds location. Cells without terminals, e.g., in China or Russia, are removed (line~10) and use no addresses. As the served footprint changes, code-table rows are appended or retired, and a region's index width $b$ can grow or shrink at the next re-planning epoch without affecting other regions.

\begin{figure}[!t]
  \centering
  \includegraphics[width=0.93\columnwidth]{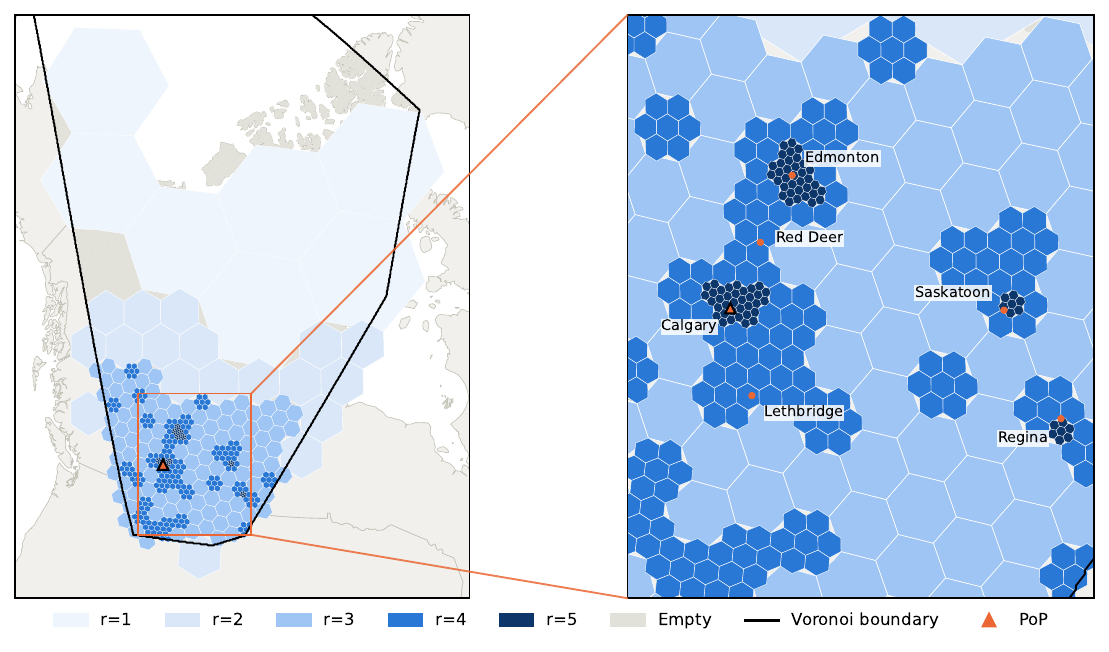}
  \vspace{-2mm}
  \caption{Compacted H3 cells, Calgary PoP region.}
  \label{fig:adaptive}
  \vspace{-3mm}
\end{figure}

\textbf{Example: the Calgary PoP region.} We apply Algorithm~\ref{alg:compact} to the land of the Calgary PoP's Voronoi region (23,546 cells at $r{=}5$) with 65,178 illustrative terminals derived from population, using the /28 pool ($2^S{=}512$, $\theta{=}0.5$, $k{=}10$, $\gamma{=}0.25$). Compaction yields 423 cells that need 465 indices, at most 226 per resolution, so a 9-bit index suffices (Fig.~\ref{fig:adaptive}, Table~\ref{tab:compaction}). The 88 $r{=}5$ city cells cover 0.3\% of the land but hold 37\% of the terminals. A uniform $r{=}4$ grid would need 1,153 indices ($>2^{10}$), and a uniform $r{=}3$ grid has a 42\,km error, versus 25\,km ($^{\dagger}$demand-weighted in Table~\ref{tab:compaction}) after compaction.

\begin{algorithm}[!t]
\caption{Demand-aware compaction and indexing}
\label{alg:compact}
\footnotesize
\begin{algorithmic}[1]
\Require cells at $r{=}5$ with $d_c$; $2^S$, $\theta$, $k$, $\gamma$, $r_{\mathrm{pol}}$
\State $\mathcal{C}\gets$ land cells at $r{=}5$
\For{$r = 5$ \textbf{down to} $2$}
  \For{each parent $q$ at $r{-}1$, in H3-ID order}
    \State $D\gets\sum_{c\in\mathrm{children}(q)} d_c$ \Comment{missing children count as 0}
    \If{$D{<}\theta 2^{S} \lor \exists c{:}\,0{<}d_c{<}k \lor r{>}\min(r_{\max}(\kappa),r_{\mathrm{pol}}(q))$}
      \State $\mathcal{C}\gets\textsc{CompactCells}(\mathcal{C},q)$ \Comment{replace children by $q$}
    \EndIf
  \EndFor
\EndFor
\State remove cells with $d_c=0$ \Comment{added when a terminal appears}
\For{each $r\in\{1,\dots,5\}$: $i\gets0$; cells $c$ by $n_c$ (Eq.~\ref{eq:indices}) desc., H3 ID}
  \State $i\gets$ next multiple of $n_c$; assign $[i,i{+}n_c)$ to $c$; $i\gets i{+}n_c$
\EndFor
\State \textbf{if} some $i>2^{b}$ \textbf{then} increase $\theta$ and repeat
\end{algorithmic}
\end{algorithm}

\subsection{GeoIP Publishing and Supporting Mobility}
\label{sec:encoding}

Starlink can publish one \emph{code table} per prefix (Table~\ref{tab:codetable}). Each row maps a block of indices at one resolution to one H3 cell and its /56 range; a GeoIP consumer decodes an address by matching its prefix and reading the 3-bit resolution and the 9-bit index, and finding the row whose block contains it (R3). Unlike the GeoFeed, the table describes the grid, not users. Rows are append-only: a growing group first receives additional indices and is split only at planned re-planning epochs.
\par When a terminal changes cell, its Locator changes while its internal Endpoint Identifier stays stable. Mobile terminals use coarser cells ($r\le r_{\max}(\kappa)$), and ships and aircraft use $r{=}1$--2, which also covers oceans. Hysteresis avoids flapping at cell boundaries, and make-before-break keeps the old /56 valid briefly.

\begin{table}[!t]
\centering
\caption{StarLoc code table of Calgary (\texttt{2605:59c1:8000::/35}).}
\label{tab:codetable}
\vspace{-1mm}
\scriptsize
\setlength{\tabcolsep}{2.6pt}
\begin{tabular}{@{}c r r l l@{}}
\toprule
\textbf{res} & \textbf{start} & $n$ & \textbf{H3 cell} & \textbf{Prefix} \\
\midrule
5 & 000 & 4 & \texttt{8512eccbfffffff} (Edmonton) & \texttt{2605:59c1:9400::/45} \\
5 & 014 & 2 & \texttt{8512ccd7fffffff} (Calgary) & \texttt{2605:59c1:941c::/46} \\
4 & 020 & 2 & \texttt{8412ea9ffffffff} (Red Deer) & \texttt{2605:59c1:9028::/46} \\
1 & 006 & 1 & \texttt{81137ffffffffff} (Yellowknife) & \texttt{2605:59c1:840c::/47} \\
\bottomrule
\end{tabular}
\vspace{-3mm}
\end{table}

\begin{table}[!t]
\centering
\caption{Compaction of the Calgary region.}
\label{tab:compaction}
\vspace{-1mm}
\scriptsize
\setlength{\tabcolsep}{2.8pt}
\begin{tabular}{@{}c r r r r r r@{}}
\toprule
$r$ & \textbf{Cells} & \textbf{Area (km$^2$)} & \textbf{Terminals} & \textbf{Indices} & \textbf{Term./cell} & $E_R$ \\
\midrule
1 & 7 & 4,497,336 & 652 (1\%) & 7 & 26--243 & 295\,km \\
2 & 18 & 1,612,983 & 2,373 (4\%) & 20 & 1--551 & 111\,km \\
3 & 103 & 1,261,461 & 15,955 (24\%) & 104 & 15--503 & 42\,km \\
4 & 207 & 356,675 & 22,397 (34\%) & 226 & 12--795 & 16\,km \\
5 & 88 & 21,865 & 23,801 (37\%) & 108 & 21--897 & 6\,km \\
\midrule
Total & 423 & 7,750,321 & 65,178 & 465 & & 25\,km$^{\dagger}$ \\
\bottomrule
\end{tabular}
\vspace{-3mm}
\end{table}

\section{Evaluation} \label{sec:eval}

We evaluate StarLoc's geolocation accuracy against current
Starlink PoP-based geolocation. Since LISP, HIP, and ILNP do not encode geographic location, PoP-based positioning
serves as the topological baseline. We use two metrics: mean positioning error and geographic
index bits. For $N$ sampled locations in PoP region $p$,
the mean error is
\begin{equation}
\bar{E}_p(r)=\frac{1}{N}\sum_{i=1}^{N}
d_{\mathrm{gc}}(x_i,\hat{x}_i(r)),
\end{equation}
where $d_{\mathrm{gc}}$ is the great-circle distance between
actual location $x_i$ and its estimated location
$\hat{x}_i(r)$, either the PoP or H3 cell center.
A uniform grid with $C_{p,r}$ cells requires
$b=\lceil\log_2 C_{p,r}\rceil$ index bits. After compaction,
$b=\lceil\log_2\max_r I_{p,r}\rceil$, where $I_{p,r}$
counts allocated index positions at resolution $r$,
including multiple indices for high-demand cells.
The total Locator width is $L=3+b$, including
3 resolution bits.

\begin{figure}[!t]
  \centering
  \includegraphics[width=0.8\columnwidth]{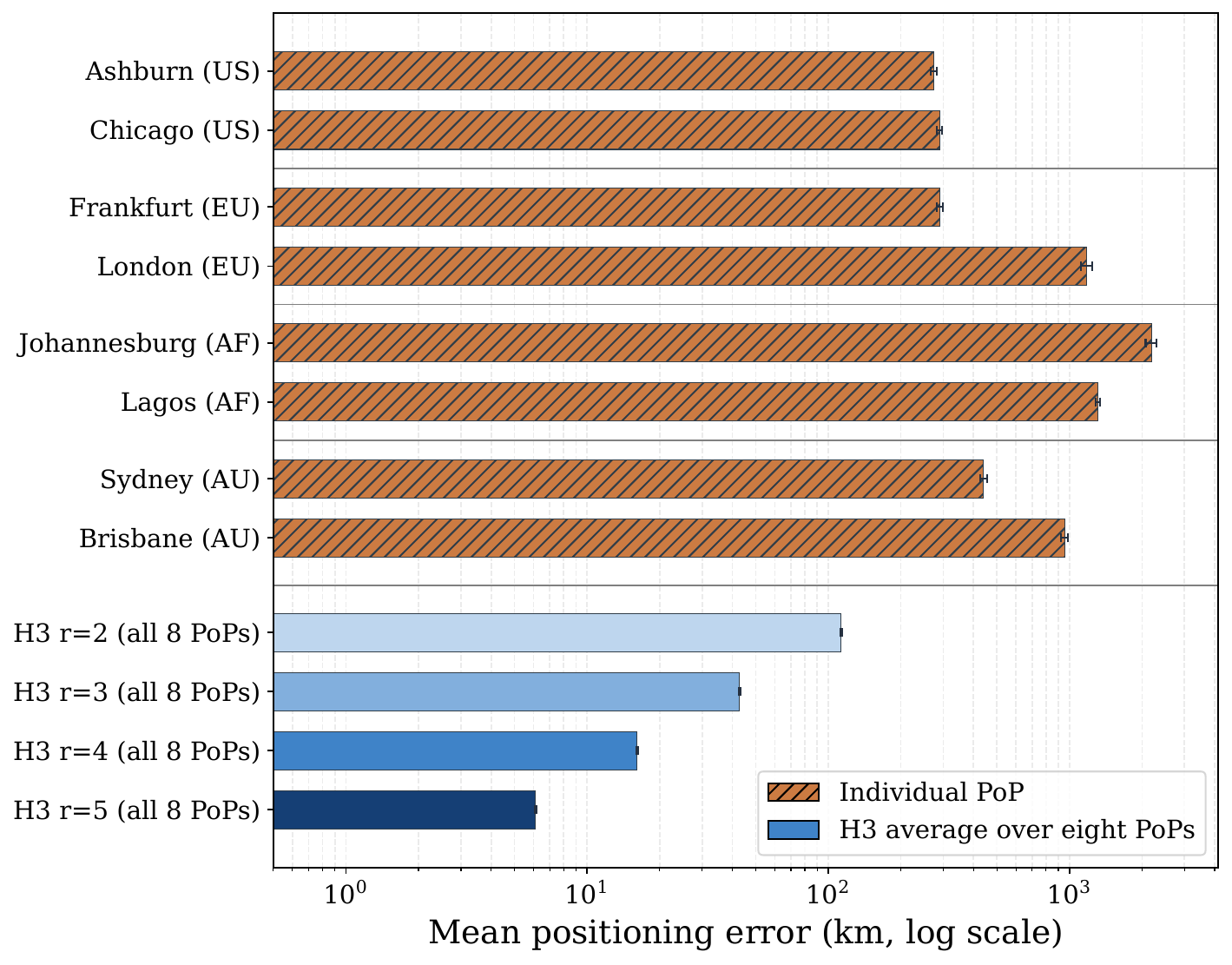}
  
  \caption{Mean positioning error of PoP-based location
  vs.\ H3 Locators at $r=2$--$5$.}
  \label{fig:pop_vs_h3}
\end{figure}

Fig.~\ref{fig:pop_vs_h3} compares PoP-based and H3
positioning errors using 1,000 sampled land locations
per PoP across eight PoPs on four continents, with
95\% confidence intervals from 2,000 bootstrap resamples.
PoP errors range from 275\,km (Ashburn) to 2,203\,km
(Johannesburg), whereas H3 achieves nearly
region-independent mean errors of 113, 43, 16, and
6.1\,km at $r=2$--5, respectively. At $r=5$,
H3 reduces the average error by approximately
141$\times$ compared with PoP-based positioning.

\begin{table}[!t]
\centering
\caption{Active /56s (2026-10-18) and number
of H3 land cells per resolution in the selected PoP regions.}
\label{tab:cells_bits}
\vspace{-1mm}
\footnotesize
\setlength{\tabcolsep}{4pt}
\begin{tabular}{@{}l r r r r r r@{}}
\toprule
\textbf{PoP region} & \textbf{Active IPv6} &
$r=1$ & $r=2$ & $r=3$ & $r=4$ & $r=5$ \\
\midrule
Ashburn & 309,066 & 3 & 9 & 44 & 248 & 1,516 \\
Chicago & 334,246 & 3 & 13 & 57 & 334 & 2,099 \\
Frankfurt & 275,800 & 4 & 13 & 65 & 369 & 2,329 \\
London & 269,284 & 10 & 35 & 138 & 661 & 3,527 \\
Johannesburg & 178,415 & 26 & 91 & 414 & 2,427 & 15,907 \\
Lagos & 78,162 & 25 & 107 & 603 & 3,878 & 26,184 \\
Sydney & 183,393 & 4 & 13 & 50 & 258 & 1,607 \\
Brisbane & 116,266 & 21 & 55 & 188 & 905 & 5,339 \\
\bottomrule
\end{tabular}
\vspace{-3mm}
\end{table}

Table~\ref{tab:cells_bits} shows that cell counts
reflect geographic area rather than user demand.
Johannesburg and Lagos contain 72\% of all $r=5$
cells but only 15\% of active IPv6 addresses.
Lagos requires 26,184 cells versus 1,516 in Ashburn,
despite having four times fewer active addresses.
A uniform $r=5$ grid requires 15 index bits in Lagos,
while 9 bits support only 512 cells. This motivates
demand-aware resolution instead of uniform cell allocation.

\begin{figure}[!t]
  \centering
  \begin{subfigure}[b]{0.8\columnwidth}
    \centering
    \includegraphics[width=\columnwidth]{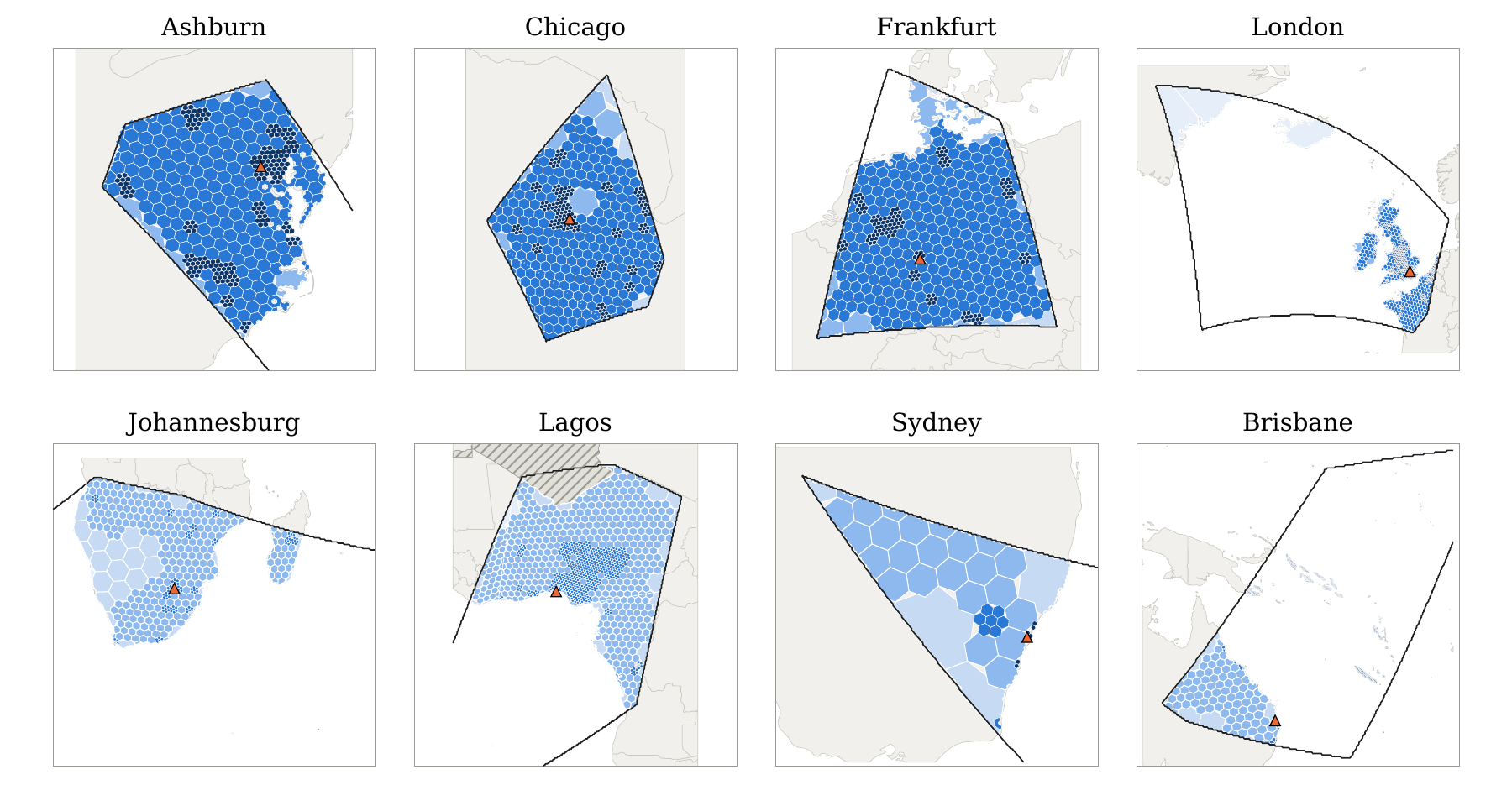}
    \vspace{-5mm}
    \caption{Compacted cells per PoP region.}
    \label{fig:pop_maps}
  \end{subfigure}
  \vspace{1mm}
  \begin{subfigure}[b]{0.8\columnwidth}
    \centering
    \includegraphics[width=\columnwidth]{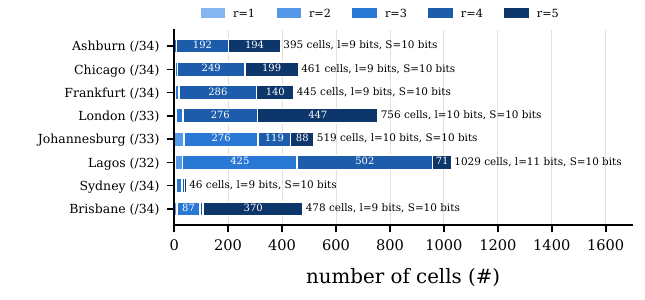}
    \vspace{-5mm}
    \caption{Cells per resolution, prefix, and customer slot bits.}
    \label{fig:cell_mix}
  \end{subfigure}
  \vspace{-2mm}
  \caption{StarLoc compaction in eight PoP regions.}
  \label{fig:pop_compaction}
  \vspace{-3mm}
\end{figure}

Fig.~\ref{fig:pop_compaction} shows demand-aware H3
compaction, retaining fine $r=5$ cells in populated
areas (194 around Washington--Baltimore and 447 in
London) while using coarse $r=2$--3 cells in sparse
regions such as Namibia and inland Australia.
With $S=10$ (1,024 /56s per index), most regions
require 9 index bits (/34), while London and
Johannesburg require 10 (/33) and Lagos 11 (/32).
Thus, StarLoc represents 46--1,029 compacted cells
using only 12--14 Locator bits, balancing geographic
accuracy and address-space efficiency.

\section{Conclusion} \label{sec:concl}

In this paper, we analyzed Starlink's IPv6 addressing
and identified limitations in geographic accuracy,
prefix allocation, and GeoIP stability. To address
these challenges, we proposed StarLoc, which encodes
physical location into IPv6 using adaptive H3 cells
while preserving /56 delegations and separating
geographic Locators from stable Endpoint Identifiers.
Our evaluation shows that StarLoc reduces mean
positioning error by up to 141$\times$ compared with
PoP-based geolocation, while demand-aware cell
compaction limits Locator overhead to 12--14 bits.
These results demonstrate the feasibility of accurate,
compact, and mobility-aware geographic addressing
for LEO satellite networks.

\bibliographystyle{IEEEtran}
\bibliography{software}

\end{document}